\documentclass[astra]{copernicus}
\begin{document}

\title{Enhancement of the Yakutsk array by atmospheric Cherenkov telescopes to study cosmic rays above $10^{15}$ eV}

\author{A. A. Ivanov, S. P. Knurenko, Z. E. Petrov, M. I. Pravdin, and I. Ye. Sleptsov}
\affil{Shafer Institute for Cosmophysical Research \& Aeronomy, Yakutsk 677980, Russia}

\runningtitle{Enhancement of the Yakutsk array}

\runningauthor{A. A. Ivanov et al.}

\correspondence{A. A. Ivanov\\ (ivanov@ikfia.ysn.ru)}

\firstpage{1}

\maketitle
\vspace*{-0.54cm}

\begin{abstract}
The aim of the Yakutsk array enhancement project is to create an instrument to study the highest-energy galactic cosmic rays (CRs) -- their sources, energy spectrum, and mass composition. Additionally, there will be unique capabilities for investigations in the transition region between galactic and extragalactic components of CRs.

Using the well-developed imaging atmospheric Cherenkov telescope technique adapted to the energy region $E>10^{15}$ eV, we plan to measure the longitudinal structure parameters of the shower, e.g., angular and temporal distributions of the Cherenkov signal related to $X_{max}$ and the mass composition of CRs.

The main advantages of the Yakutsk array, such as its multi-component measurements of extensive air showers, and model-independent CR energy estimation based on Cherenkov light measurements, will be inherited by the instrument to be created.
\end{abstract}

\introduction
The cosmic ray (CR) astrophysics behind extensive air shower (EAS) measurements in the energy range $E>10^{14}$ eV pose several problems of great interest that stimulate further in-depth research. In addition to general objectives such as CR energy spectrum, mass composition, and sources, some specific questions should be addressed in the immediate future. For example: What is the maximum energy of supernova remnant (SNR) accelerators? Is the 'knee' of CR spectrum due to the diffusion of particles in magnetic fields or to the upper limit of galactic sources? Where is the transition region between galactic and extragalactic components of CRs?

Some theoretical predictions may soon be verified experimentally. In particular, it is shown in nonlinear kinetic theory of CR acceleration in SNRs that the maximum energy of nuclei is proportional to their charge \citep{Shock}. Thus, the knee in the energy spectrum is the result of the contribution of progressively heavier species, and galactic CRs are dominated by iron-group nuclei at $E\sim10^{17}$ eV. An alternative model is proposed by \citet{SSM} in which a recent nearby SNR accelerates CRs resulting in rather different structure in the energy spectrum, detectable as well.

As an initial step in enhancement of CR detecting possibilities, the Yakutsk array was recently equipped with a pinhole camera with a row of photomultiplier tubes (PMTs) in the rear as a prototype detector of the angular distribution of Cherenkov light induced by showers in the atmosphere \citep{PinHole}. A conclusion from the measurements was that angular and temporal characteristics of the signal can be used to investigate longitudinal shower profile in the atmosphere.

Developing these efforts further, we plan to use new device consisting of multi-anode PMT at the focus of a spherical mirror as a Cherenkov light detector embedded in the array.

\begin{figure}[t]
\center\includegraphics[width=0.75\columnwidth]{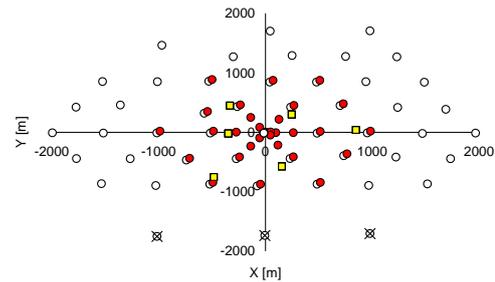}
\caption{Plan of the Yakutsk array. Captions: open circles = soft component detectors; filled circles = Cherenkov light detectors; squares = muon detectors. Three stations at the bottom (indicated by crosses) were destroyed by a flood in 2010.}
\label{fig:Map}
\end{figure}

\section{Present status of the Yakutsk array, and the next-step astrophysical goals}
At present, the Yakutsk array is measuring the soft component of EAS with 58 stations, muons with 6 underground detectors, and Cherenkov light with 48 PMTs. All detectors are irregularly distributed within a 10 km$^2$ array area (Fig.~\ref{fig:Map}); the target energy range of investigations is $10^{15}$ eV to $10^{20}$ eV. More technical details and physical results are given in \citet{Mono,CRIS,NJP,APJ} and \citet{site}.

Our next task is to modernize the array in order to obtain a precise instrument capable of measuring the highest-energy galactic CRs -- their sources, energy spectrum, and mass composition. Related to this, we also aim to study a transition region between galactic and extragalactic components of CRs where some irregularities in spectrum and composition may be revealed.

A crucial clue in this approach is the accurate determination of the mass composition of CRs, which is a weak point of existing giant EAS arrays. In this context we plan to adapt the well-known atmospheric Cherenkov telescope technique \citep[e.g.,][]{IACT} to measure the angular and temporal structure of the signal connected to EAS longitudinal profiles above $E=10^{15}$ eV.

The idea is not to concentrate on discriminating the gamma-ray initiated showers from the hadronic showers; instead, all showers from different primary particles will be detected by the wide field of view (FoV) telescopes in coincidence with the array detectors. The angular and temporal parameters of showers measured will then be analyzed to identify EAS primary particles. Longitudinal shower profile measurements can also be used to estimate some hadronic interaction parameters \citep{WeiHai}.

Experimental arguments in elucidating the origin of the knee and ankle in CR spectrum will significantly strengthen due to the estimation of mass composition in the energy range relevant. Existing models (e.g., listed in
Introduction) are very different in composition expected around the knee and ankle, so the estimation of the average mass of the primaries in addition to the improved measurement of the sharpness of the knee/ankle should allow us to discriminate between models.

Another important step in the adaptation is the essential reduction in the size and cost of the telescope. This is possible through increased threshold energy because the total number of Cherenkov photons emitted by EAS relativistic electrons is proportional to the shower energy. If we compare one of the High-Energy Stereoscopic System (HESS) telescopes (diameter $D=13$ m, $E_{thr}=10^{11}$ eV /www.mpi-hd.mpg.de/hfm/HESS/HESS.shtml/) with that reduced to $D=13$ cm, then the number of Cherenkov photons detected from EAS is comparable at $E>10^{15}$ eV. Hence, in our energy range, the set of Cherenkov telescopes with $D=13$ cm is approximately equivalent to HESS functioning in the energy range $E>10^{11}$ eV; this is, of course, excepting the event rate: CR intensity ratio in two energy intervals is $J(E>10^{15})/J(E>10^{11})=10^{-4\kappa}=1.6\times10^{-7}$, where $\kappa\simeq1.7$ is the integral spectrum index.

\section{Modeling the measurement of EAS longitudinal structure parameters with Cherenkov telescopes}
It is well known that the angular and temporal structure of Cherenkov light emitted by EAS can be used to infer the longitudinal development parameters of a shower \citep{Zatsepin, Fomin, Prosin}. The angular distribution of a Cherenkov signal from EAS was calculated by \citet{Zatsepin}, where it was assumed that the angular distribution of emitted photons is determined by that of electrons in the shower. \citet{Fomin} then proposed using the pulse shape of the Cherenkov signal, namely the pulse width, to indicate the shower maximum position $X_{max}$ in the atmosphere.

Experimental measurements of the Cherenkov signal pulse shape were performed in Yakutsk in 1973--1979. The pulse width was measured in the core distance interval (100,800) m for the shower size $1.9\times10^7<N<1.3\times10^8$ \citep{Mono}. The results were used not only to estimate $X_{max}$, but attempts were also made to evaluate the EAS cascade curve at energies around $10^{17}$ eV \citep{Prosin,Knrnk}. Unfortunately, the angular and temporal resolutions of optical instruments and data-acquisition systems at that time were not sufficient to obtain the accuracy required to distinguish EAS primaries and/or nuclear interaction models.

To verify the ability of the Cherenkov telescope to discriminate angular and temporal profiles of signals from
showers initiated by primary nuclei and photons at energies above $10^{15}$ eV, we modeled the
process. The same assumptions as in \citet{Mono} are used to describe the Cherenkov light emission in atmosphere by  the vertical EAS: The energy spectrum of electrons is taken at $s=1$ (shower maximum); the spatial distribution of
electrons is neglected; angular distributions of electrons and Cherenkov photons are Gaussian; and the time of flight to  detector for photons is given by $t=\frac{n}{c}\sqrt{h^2+R^2}-h/c$, where $n$ is the refraction coefficient of air, $h$  is height, and $R$ is axis distance.

In this crude approximation the number of photons is given by the integral of the EAS cascade curve multiplied by the function giving the contribution of an electron to the Cherenkov light flux at a given $R$ and depth in the atmosphere \citep{Mono}. As the cascade curve, we used the gamma-distribution approximation of the HiRes results \citep{HiRes}. In the energy range $(10^{15},10^{17})$ eV beyond the reach of HiRes, we used neXus2 model simulation results for showers initiated by $P, Fe$ \citep{neXus}. A difference in $X_{max}(E)$ of showers from proton and iron nuclei in this model was also employed as the input parameter for the cascade curve from HiRes. For the $\gamma$-initiated showers, the Greisen formula was used.

\begin{figure}[t]
\center\includegraphics[width=0.75\columnwidth]{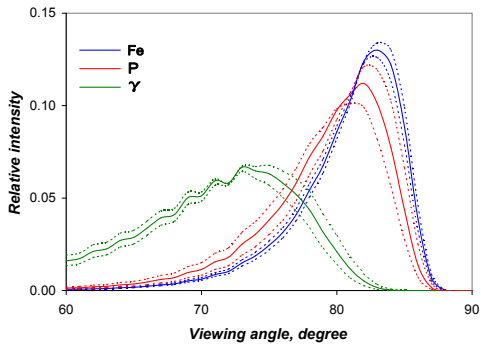}
\caption{Angular spread of EAS Cherenkov light in the detector at $R=800$ m from the axis. Vertical
showers ($E=10^{15}$ eV) are shown. The areas within dotted curves represent the intrinsic fluctuations of the showers.}
\label{fig:Angular}
\end{figure}

In Fig.~\ref{fig:Angular} the angular distribution of Cherenkov light in the detector at $R=800$ m is shown. The EAS primaries are $P, Fe$, and $\gamma$ with $E=10^{15}$ eV. It appears from this that the angular resolution $\sim 1^0$ of the Cherenkov telescope is sufficient to distinguish the primaries ($E>10^{15}$ eV), at least the $\gamma$-quanta from the nuclei, based on the angular distribution of the signal alone.

\begin{figure}[t]
\center\includegraphics[width=0.75\columnwidth]{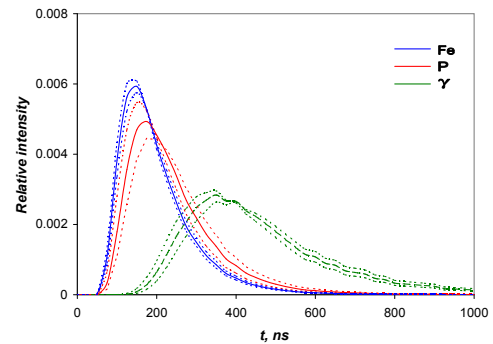}
\caption{Temporal structure of the Cherenkov signal from the vertical EAS, $E=10^{15}$ eV, $R=800$ m, induced by the primary particles indicated. Symbols are the same as in the previous Figure.}
\label{fig:Temporal}
\end{figure}

Another parameter that has been suggested by the pioneers is the temporal characteristic of the Cherenkov signal from EAS. In Fig.~\ref{fig:Temporal} the results of our simulations are shown. In this case the approximation used is too rough, so that the half width of the signal is overestimated by $\sim40\%$ in comparison with the actual value measured in Yakutsk \citep{Mono} and with more precise simulations \citep[e.g.,][]{Klmkv}. Nevertheless, assuming that the overestimation is independent of the primary particle type (diminishing the widths of the signals to 0.7 in all three cases), we can conclude that the time resolution $\delta t<10$ ns is sufficient to distinguish the primaries at $E>10^{15}$ eV.

Shower fluctuations characterized by $\sigma_{Xmax}=67/22/47$ g/cm$^2$, for $P/Fe/\gamma$ primaries respectively, are shown by the dotted curves in Figs \ref{fig:Angular},~\ref{fig:Temporal}. Conclusions about the possibility to distinguish primaries do not weaken because of fluctuations less than the differences caused by the primaries. More reliable results will be achieved by combining both the angular and the temporal measurements in one experiment.

\section{Preliminary Cherenkov telescope design, and enhancement of the Yakutsk array}
Recent developments in the PMT industry have made it possible to dramatically reduce the size of Cherenkov telescopes. For example, the imaging camera for the HESS telescope consists of 960 PMTs each with a photocathode area of $\sim16$ cm$^2$ (pixel size). Proportionally reducing the area of mirror and camera 10$^4$ times, we should have a pixel size of $\sim0.4\times0.4$ mm$^2$ for the same angular resolution.

A newly developed semiconductor light-sensor, the so-called Geiger-mode avalanche photodiode (G-APD, or 'silicon PMT'), typically has a sensitive area of $\sim1\times 1$ mm$^2$ covered by 100 to 10$^4$ cells. This is larger than we need, but on the other hand, the mirror diameter can be greater than 13 cm. Application of this type of light-sensor is currently being widely discussed -- see for example \citet{SiPM} -- and there are examples of cameras designed with appropriate data-acquisition systems \citep[e.g.,][]{DRS}.

Another type of light-sensor applicable to the Cherenkov mini-telescope is position-sensitive PMT. In this case the flat multi-anode is segmented in $n\times n$ pixels ($n=2,..,16$ for Hamamatsu products; pixel area $\geq2\times 2$ mm$^2$).

We selected Hamamatsu R2486 PMT with a $16\times 16$ crossed wire anode for the prototype detector. The PMT is used as an imaging camera placed near the focus of a spherical mirror. Because the radius of the R2486 photocathode is 3 cm, the mirror diameter was chosen to be D = 26 cm, and the radius of curvature R = 22.5 cm. The mirror parameters were optimized to provide as wide a FoV as possible where the image distortions are less than the given pixel size.

Angular resolution of the prototype telescope is determined by the pixel size $d=3.8$ mm and focal distance $\delta\alpha=2d/R\simeq2^0$. The width of FoV was estimated as $(-14^0,14^0)$ in regard to the optical axis.

\begin{figure}
\center\includegraphics[width=0.75\columnwidth]{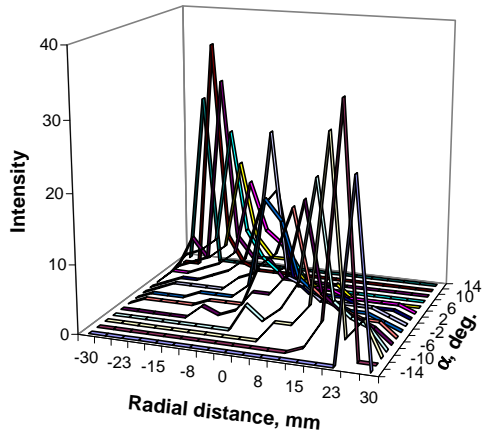}
\caption{Scanning angular distribution of light from a distant point source at the slant angle $\alpha$ with the position-sensitive PMT in the focus of the spherical mirror.}
\label{fig:Scans}
\end{figure}

The result of modeling in the plane intersecting the mirror optical axis is shown in Fig.~\ref{fig:Scans}. The light intensity emitted by the distant point sources with slant angles $-14^0\leq\alpha\leq14^0$ is shown as a function of the axis distance on the camera surface. The root-mean-square radius of the light spot averaged over camera surface is 0.64 of the pixel size. Thus, the position of intensity maximum is an explicit function of incident angle. This demonstrates the ability of our prototype Cherenkov telescope to convert the angular distribution of light sources to the spatial distribution of signals on the imaging camera surface with adequate accuracy.

The effective acceptance area of our prototype telescope is $\sim200$ cm$^2$ taking into account the device geometry, shadowing of the mirror, etc. This is comparable to the area of FEU-49 PMT (176 cm$^2$) in use in Cherenkov detectors of the Yakutsk array. Thus, the real EAS event rate previously detected with the Cherenkov sub-array can be used as an estimator of the expected number of showers. In the arrangement of Cherenkov light detectors shown in Fig.~\ref{fig:Map}, the number of showers detected per hour at $E>1.2\times10^{15}$ eV is 62, and 4 events were observed at $E>6\times10^{16}$ eV \citep{CERN}.

If we place the set of Cherenkov telescopes in a congruent hierarchical triangle structure as in Fig.~\ref{fig:Map} (filled circles), then the expected number of EASs detected per year (assuming 500 hours/year is suitable for Cherenkov light measurements) is $\sim31000$ at $E>1.2\times10^{15}$ eV, and $\sim2000$ at $E>6\times10^{16}$ eV.

\conclusions
We have demonstrated the possibility of investigating EAS longitudinal development in the atmosphere through measuring the angular and temporal distributions of a Cherenkov signal with a telescope embedded in a surface EAS array environment. The enhancement of the Yakutsk array by a set of Cherenkov telescopes adapted to the energy range $E>10^{15}$ eV is aimed at significantly improving the array's ability to study the mass composition of CRs. The main advantage of this combination consists in the integration of the multi-component measurement of EAS and model-independent CR energy estimation by the surface array with excellent angular and temporal resolution of the Cherenkov telescopes.

\begin{acknowledgements}
The work is supported in part by SB RAS (integral project 'Modernization of the Yakutsk array'), RFBR (grant no. 09-02-12028) and Russian ministry of education and science (contract nos. 02.518.11.7173, 02.740.11.0248).
\end{acknowledgements}


\begin{thebibliography}{}

\bibitem[Abu-Zayyad et al.(2001)]{HiRes} Abu-Zayyad, T. et al.: A measurement of the average longitudinal development profile of CR air showers between $10^{17}$ eV and $10^{18}$ eV, Astropart. Phys., 16, 1--11, 2001.

\bibitem[Anterhub et al.(2009)]{DRS} Anterhub, H. et al.: A novel camera type for VHE gamma-ray astronomy based on G-APD, astro-ph/0911.4920, 2009.

\bibitem[Berezhko and V\"{o}lk(2007)]{Shock} Berezhko, E. G. and V\"{o}lk, H. J.: Spectrum of CRs produced in SNRs, astro-ph/0704.1715, 2007.

\bibitem[Dyakonov et al.(1991)]{Mono} Dyakonov, M. N., Efimov, N. N., Egorov, T. A. et al.: Cosmic rays of extremely high energy, Nauka, Novosibirsk, USSR, 251 pp., 1991 (in Russian).

\bibitem[Egorova et al.(2004)]{CRIS} Egorova, V. P., Glushkov, A. V., Ivanov, A. A. et al.: The spectrum features of UHECRs below and surrounding GZK, Nucl. Phys. B (Proc. Suppl.), 136, 3--11, 2004.

\bibitem[Erlykin and Wolfendale(2001)]{SSM} Erlykin, A. D. and Wolfendale, A. W.: Structure in the cosmic ray spectrum: an update, J. Phys. G: Nucl. Part. Phys., 27, 1005-1030, 2001.

\bibitem[Fomin and Khristiansen(1971)]{Fomin} Fomin, Yu. A. and Khristiansen, G. B.: Yader. Fiz., 14, 642, 1971.

\bibitem[Garipov et al.(2001)]{PinHole} Garipov, G. K. et al.: The Cherenkov track detector consisting of the Yakutsk array, in: Proceedings of the 27th International Cosmic Ray Conference, Hamburg, Germany, 3, 885--888, 2001.

\bibitem[Grigorjev et. al.(1979)]{Prosin} Grigorjev, V. M. et al.: Recovering EAS cascade curve from the Cherenkov pulse shape at energy $E>10^{17}$ eV, JETP Lett., 30, 747--750, 1979.

\bibitem[Hinton(2009)]{IACT} Hinton, J.: Ground based gamma-ray astronomy with Cherenkov telescopes, New Journ. Phys., 11, 055005 (1--17), 2009.

\bibitem[Ivanov et al.(2003)]{CERN} Ivanov, A. A., Knurenko, S. P., and Sleptsov, I. Ye.: The energy spectrum of CRs above $10^{15}$ eV derived from air Cherenkov light measurements in Yakutsk, Nucl. Phys. B (Proc. Suppl.), 122, 226--230, 2003.

\bibitem[Ivanov et al.(2009)]{NJP} Ivanov, A. A., Knurenko, S. P., and Sleptsov, I. Ye.: Measuring EAS with Cherenkov light detectors of the Yakutsk array: The energy spectrum of CRs, New Journ. Phys., 11, 065008, (1--30), 2009.

\bibitem[Ivanov(2010)]{APJ} Ivanov, A. A.: Comparing the energy spectra of UHECRs measured with EAS arrays, Astrophys. Journ., 712, 746-751, 2010.

\bibitem[Kalmykov et al.(1975)]{Klmkv} Kalmykov, N. N. et al.: Investigation of the pulse shape of EAS Cherenkov radiation, JETP Lett., 21, 66--70, 1975.

\bibitem[Knapp et al.(2001)]{neXus} Knapp, J. et al.: Extensive air shower simulations at the highest energies, Astropart. Phys., 19, 77--99, 2003.

\bibitem[Knurenko et al.(2001)]{Knrnk} Knurenko, S. P. et al.: Cherenkov radiation of CR EAS. Part 3. Longitudinal development of showers in the energy region of $10^{15}-10^{17}$ eV, in Proceedings of the 27th International Cosmic Ray Conference, Hamburg, Germany, 1, 157--160, 2001.

\bibitem[Knurenko et al.(2008)]{WeiHai} Knurenko, S. P. et al.: Recent results from the Yakutsk experiment: The development of EAS and the energy spectrum and primary particle mass composition in the energy region of $10^{15}-10^{19}$ eV, Nucl. Phys. B (Proc. Suppl.), 175-176, 201--206, 2008.

\bibitem[Teshima et al.(2007)]{SiPM} Teshima, M. et al.: SiPM development for astroparticle physics applications, in: Proceedings of the 30th International Cosmic Ray Conference, Merida, Mexico, 2, 985--988, 2007.

\bibitem[the website(2006)]{site} Website of the Yakutsk array: http://eas.ysn.ru, 2006.

\bibitem[V. Zatsepin(1964)]{Zatsepin} Zatsepin, V. I.: Angular distribution of the Cherenkov light intensity induced by EAS of CRs, JETP, 47, 689-695, 1964.

\end{thebibliography}
\end{document}